***Beltanelliformis brunsae*** **Menner in Keller et al., 1974: an undoubted Ediacaran fossil from Neoproterozoic of Dobrogea (Romania).**

Jean-Paul SAINT MARTIN & Simona SAINT MARTIN

UMR 7207 CR2P MNHN/CNRS/UPMC/Sorbonne Université, Centre de Recherches sur la Paléobiodiversité et les Paléoenvironnements, Muséum national d'Histoire naturelle, département Origines et Evolution, 8 rue Buffon, 75231 Paris cedex 05, France. jpsmart@mnhn.fr; simsmart@mnhn.fr

Corresponding author : Jean-Paul Saint Martin ; jpsmart@mnhn.fr

**ABSTRACT**

Ediacaran fossils are now largely known in different parts of the world. However, some countries are poorly documented on these remains of a still enigmatic life. Thus, rare fossils from the Neoproterozoic Histria Formation of central Dobrogea (Romania) have been reported. Two specimens with discoid imprints are described here in detail and assigned to the typical Ediacaran species *Beltanelliformis brunsae* Menner in Keller et al., 1974. This paleontological development confirms both the large geographical distribution of this species and the Ediacaran age of the Histria Formation.

**RÉSUMÉ**

Les fossiles édiacariens sont maintenant largement connus à travers le monde. Cependnant, certaines régions restent peu documentées sur ces restes, témoignages d'une vie qui apparaît encore aujourd'hui assez énigmatique. Ainsi, seuls de rares fossiles provenant de la Formation Histria du Néoprotérozoïque de Dobrogea centrale (Roumanie) avaient été signalés. Deux échantillons portant des impressions discoïdes sont ici décrits en détail et attribués à l'espèce typiquement édiacarienne *Beltanelliformis brunsae* Menner in Keller et al., 1974. Cette identification permet de confirmer la large répartition de cette espèce et l'âge édiacarien de la Formation Histria.

**MOTS-CLÉS**

Empreintes discoïdes plates, Biota édiacarien, Précambrien, Dobrogea, Roumanie.

**INTRODUCTION**

The period corresponding to the Late Precambrian (Neoproterozoic) and the Precambrian-Cambrian boundary is a key moment in the evolution of the biosphere and biodiversity. At this time, the metazoan organisms begin to organize in complex aquatic ecosystems prefiguring the "Cambrian revolution" from which the modern living aquatic world settles durably with its multiple forms of life (Erwin & Tweedt 2011). The Late Proterozoic fossiliferous deposits distributed throughout the world have brought incomparable information, the most famous being that of Ediacara in Australia, which gave its name to the last period of the Precambrian, the Ediacaran (-630 to -542 million years). Since the highlighting of this very particular stand in the middle of the 20th century, a particular interest has been focused on these fascinating archaic forms of life, with a somewhat idealized vision of a world without predators relayed through the notion of "Garden of Ediacara" (McMenamin 1986). Despite decades of discoveries and research, Ediacaran life remains in many respects somewhat enigmatic, both in terms of comparison with large taxonomic groups of current organisms and the ecology of these early complex ecosystems (Brasier & Antcliffe 2004; McCall 2006; McGabhann, 2014). Numerous studies have thus been devoted to progressively updated organisms, and contradictory discussions have been initiated on their biological affinity and way of life. For a long time, the "Ediacaran fauna" was naturally mentioned, the authors assimilating Ediacaran organisms to animal groups such as annelids, echinoderms, arthropods... More recently, other ways have been followed, interpreting certain Ediacaran organisms as protists (Zhuravlev 1993), as microbial colonies (Steiner, 1996; Grazhdankin & Gerdes 2007), or even very controversially as fungi or lichens (Retallack 1994; Peterson *et al.* 2003; Handcliffe & Hancy 2013; Retallak 2013a, 2013b). It has also been proposed the name of "vendozoa or vendobionta" (Seilacher, 1989; Seilacher, 1992) for these organisms that could constitute an apart "phylum", today without descendants. Abundant works were dedicated to the identification and creation of a large number of taxa giving an idea of a fairly large biodiversity of the Ediacaran biota. Moreover, new methods, such as surface scanners, modeling, geochemistry, cladistic phylogeny are being used to better understand the characteristics of certain Ediacaran fossils (Antcliffe & Brasier 2008; Laflamme & Narbonne 2008; Brasier & Antcliffe 2009; Laflamme *et al.* 2009; Brasier *et al.* 2012; Bykova *et al.* 2017; Dececchi *et al.* 2017; Cui *et al.*,2016; Schiffbauer *et al.* 2017).

Recent works often take a more detailed look at the conditions of fossilization and burial (taphonomy) which make it possible to explain both the type of preservation and also to specify the original morphology of organisms (Tarhan 2010; Tarhan *et al.* 2010; Elliott *et al.* 2011). Thus, the intervention of microbial mats in the quality of preservation of Ediacaran organisms is put forward by many authors (Callow & Brasier 2009; Gehling & Droser 2009; McIlroy *et al.* 2009; Laflamme *et al.* 2010; Kenchington & Wilby 2014; Tarhan *et al.* 2017). Nevertheless, the interest of scientific community is naturally focused on the Ediacaran "system" with a particular regard on its palaeoecological significance in the ancient biosphere and its place in the general evolution of life (Xiao & Laflamme 2008; Peterson *et al.* 2008; Liu *et al.* 2010; Retallack 2010; Serezhnikova 2010; Shen *et al.* 2008; Grazhdankin 2014; Narbonne *et al.* 2014; Antcliffe 2015; Droser & Gehling 2015; Liu *et al.* 2015a; Tarhan & Laflamme 2015; Dufour & McIlroy 2016; Retallack 2016; Bowyer *et al.* 2017; Briggs 2017; Budd & Jensen 2017; Reid *et al.* 2017).

Many deposits have been identified throughout the world enriching the debate, but without really solving the enigmatic aspect of this life so little related to the modern world that will appear very soon after Precambrian-Cambrian boundary (Shu, 2008; Erwin & Tweedt 2011). Considering this wide range of opinions, any new discovery obviously provides essential data for the understanding of this disappeared ecosystem. The comprehensive reviews of Ediacaran deposits around the world provide an interesting assessment of the state of knowledge (McCall 2006; Fedonkin *et al*. 2007). However, in Europe, apart from the large outcrops of White Sea in Russia already known, some deposits in Ukraine, and the famous Charnwood site in England, the data are much rarer. The Precambrian sediments of Dobrogea are also potentially able to contain fossil remains and to provide eventually new lighting about Ediacaran life. Several works have thus mentioned the presence in Central Dobrogea of some imprints or traces attributable to elements of the Ediacaran living world (Oaie, 1992; Oaie 1993; Oaie 1998; Oaie *et al*. 2005; Seghedi *et al*. 2005; Oaie *et al*. 2012; Seghedi 2012). This organic remains are represented by two samples: a "medusoid" imprint identified as *Nemiana simplex* Palij 1976 for one, and multiple discoid imprints of possible *Beltanelloides*, for the other. Traces of activity were also observed and considered as belong to the *Nereites* ichnofacies. Like the other Ediacaran sites the presence of microbial mat at the surface of Precambrian beds was recently highlighted (Saint Martin *et al*. 2011; Saint Martin *et al.* 2012). The discovery of *Aspidella* type disks, which may represent holdfasts of frondose organisms, has made it possible to add new information and to propose an idealized reconstitution based on the knowledge of the moment (Saint Martin *et al*. 2013). However, observations made on the available material and the previous proposed determinations should probably be revised in the light of the most recent works about Ediacaran organisms. In this sense, the present preliminary article proposes to examine a certain type of discoid imprints sampled from the Precambrian terranes of Central Dobrogea, to give update identification from comparisons with known data and to discuss more general consequences.

**GEOLOGICAL SETTING**

The studied specimens bearing discoid imprints were collected within sediments of the basement in the central Dobrogea area (Fig. 1). The central Dobrogea is characterized by large outcrops, especially in the valleys, of sediments belonging to Precambrian and more particularly to the Histria formation (Seghedi & Oaie 1995) corresponding approximately to the "greenschist Formation" denomination of ancient authors. The Histria Formation, up to 5000 m thick as estimated from geological and geophysical data, consists of two coarse members of sandstone separated by a thinner member with pelites and siltites (Seghedi & Oaie 1995; Oaie 1999). According to several works (Seghedi & Oaie 1995; Oaie 1999; Oaie *et al.* 2005) considering sedimentological, structural and mineralogical features, the Histria Formation should be accumulated in a foreland basin setting, an interpretation consistent with results of geochemical and detrital zircon distribution data (Želaźniewicz *et al.* 2001; Želaźniewicz *et al.* 2009). Mineralogical, petrographic and sedimentological data show a basin sourced by a continental margin dominated by an active volcanic arc (Oaie *et al*. 2005; Seghedi *et al*. 2005). The low-grade metamorphic ("greenschists") and weakly deformed clastic rocks of the Histria Formation were considered as flyschoid-like sediments (Kraütner *et al.*

1988) corresponding to median to distal turbiditic sequences (Oaie 1998; Oaie *et al*. 2005; Seghedi *et al*. 2005; Balintoni *et al.* 2011).

A Late Proterozoic-Early Cambrian estimated age for the sediments of Histria Formation is documented both by geochemical K/Ar datation in the order of -572 million years (Giuşcă *et al*. 1967) and palynological assemblages (Kräutner *et al*. 1988). On other hand, U/Pb ages based on detrital zircon suggest a maximum late Ediacaran depositional age (Żelaźniewicz *et al*. 2009, Balintoni *et al.* 2011). In addition, the discovery in the fine grained members of a "medusoid imprint" identified as *Nemiana simplex* Palij (Oaie 1992; Oaie 1993), a typical Ediacaran fossil, has been invoked to confirm this Neoproterozoic age. However, the characteristics of this sample do not quite match with the original conception of this species (Palij, 1976; Palij *et al*. 1979). Descriptions or discoveries of other fossil remains undoubtedly belonging to the so-called "Ediacaran Biota" would be an additional argument for confirmation of the proposed ages.

The Histria succession is very rich in sedimentary structures represented by large trains of ripples marks well preserved on some stratification planes (Oaie 1993; Oaie 1998; Oaie *et al*. 2005). The frequent presence of surfaces marked by various wrinkled structures suggests the implication of microbial mats (Saint Martin *et al*. 2011; Saint Martin *et al.* 2012) resulting in the formation of MISS (Microbially Induced Sedimentary Structures), often described elsewhere in the deposits of the upper Proterozoic, especially during the Ediacaran period (Arouri *et al*. 2000; Bouougri & Porada 2007; Lan & Chen 2013; Kumar & Ahmad 2014; Parihar *et al*. 2015; Kolesnikov *et al*. 2017; Tarhan *et al*. 2017).

## MATERIAL AND METHODS

The specimen 1, actually exposed in the gallery of the National Museum of Geology (Bucharest), was collected near the town of Gura Dobrogei, on the banks of the Casimcea River where the Histria Formation is well outcropping. Mentioned as representing a "medusoid" imprints (Oaie 1992) close to the genus *Beltanelloides* (Oaie 1993), until now this specimen has never been described, although it is an important element of appreciation of possible remains testifying an Ediacaran life in Romanian sediments.

The specimen 2, housed in the collections of the French National Museum of Natural History of Paris (MNHN.F.A68682), was sampled at outcrops of the Histria Formation on the edge of Sinoe Lake, near the antic town of Histria during a field campaign carried out as part of a bilateral research program between the MNHN and the Romanian Geoecomar Institut.

In order to appreciate the variations of imprint dimensions, diameter of round imprints or long axis of oval imprints, were measured. Considering the deformations that mostly affect the specimen 1, it was considered preferable to measure in addition the surface of each discoid imprint. The surface measurement was so performed using the appropriate functions of the open software Image J.

A section intersecting the surface bearing the footprints and the underlying sediment was made in specimen 2 in order to observe in thin section the main petrographic characteristics.

## DESCRIPTION

The specimen 1of pelitic nature is a roughly rectangular plate measuring approximately 14 cm in length and 7 cm in width (Fig. 2a). The oxidized ferruginous surface displays 37 visible discoid flat imprints, of which 22 are integrally preserved and can be measured (Fig. 2b). Each discoid individual is smooth at the center and shows very fine concentric ridges towards the periphery. Discoid individuals are distributed contiguously or very slightly apart. In some cases, some discoid imprints appear to partially cover the neighboring imprint. All the imprints are affected by a unidirectional deformation according to an elongation giving them an oval shape reflecting the posterior tectonic constraints. Other earlier deformations have undoubtedly affected the surface of the sediment: small differences in elevation are observed between individuals with stretching of the wrinkled peripheral structures (Fig. 2b). The length of the long axis of the oval imprints ranges from 1.34 to 1.94 cm with an average of 1.66 cm. These measures show overall certain homogeneity of size.

The specimen 2 constitutes a block cut according to the dominant fracturation allowing showing both the surface and an oblique section of the original sedimentation (Fig. 2c). The surface of the block, strongly ferruginized to a thickness of about 1 mm, is approximately rectangular with a length of about 20 cm over a width of 7 to 8 cm. Discoid imprints on the surface are usually contiguous and also may overlap slightly one to other. They display the same character as the first specimens with a smooth part in the center and a finely wrinkled periphery (Fig. 2d). Unlike specimen 1, the discoid imprints are very slightly deformed with a roughly circular outline. The diameter is quite heterogeneous with values distributing between 1.03 and 2.50 cm for an average of 1.55 cm. The discoid imprints itself concern only a very thin thickness (Fig. 3a). The specimen shows a succession of fine grained beds and very thin beds of coarser siliciclastic sediment. Within these sequences are individualized two bodies of one centimeter thickness with slightly coarser grains structured in micro-HCS (Fig.3b). The characteristics of the sedimentation revealed by the polished section show that, like most identical fossils around the world, discoid imprints are represented in positive hyporelief, in bed sole position. Given the outcrop conditions, which essentially show the top surfaces of the beds, it is very difficult, if not almost impossible, to observe these fossils in place. This could explain the small number of found samples.

The measurement of the surfaces shows a disparity of average size between the two samples, the individuals of the specimen 2 showing a greater heterogeneity of distribution and on average a larger surface (Fig. 4). This reflects a fairly large variability in size within a sample or between two samples.

## ASSIGNMENT AS *BELTANELLIFORMIS BRUNSAE*

The assignment of the studied samples to the Ediacaran fossils mentioned in the abundant literature dedicated to discoidal impressions comes up against the already old problems of a nomenclature mainly related in fact to taphonomic aspects. If we only refer to remains exhibiting exactly the same characteristics, namely the more or less contiguous presence of flat discoid imprints with fine concentric wrinkles at the periphery and a smooth central part, two main designations have been adopted: *Beltanelliformis brunsae* Menner, in Keller 1974 (Keller *et al*. 1974; Narbonne & Hofmann 1987; Steiner 1996; Xiao *et al*. 2002; Narbonne 2007; Ivantsov *et al*. 2014; Ivantsov 2017) or *Beltanelloides sorichevae* Sokolov 1972 (Sokolov 1976; Glaessner 1984; Sokolov & Iwanowski 1990; Fedonkin 1992; Fedonkin & Runnegar 1992; Sokolov 1997; Fedonkin & Vickers-Rich,2007; Leonov 2007a, 2007b; Leonov & Ragozina 2007; Leonov & Rud'ko 2012). However, the adoption of a systematic status is complicated for several reasons according to the authors' conception and the supposed nature of these fossils: 1) the same designation has been used for variable preservation modes; 2) different names have been assigned to the same type of fossil; (3) different names have been assigned by some authors to fairly similar discoid imprints which are supposed to be different in nature but corresponding for other authors to the same original type of organism. In general, these are rather gregarious forms preserved in a bug-shaped manner or flat imprints, assuming originally a rather globular form. Narbonne & Hofmann (1987) had already distinguished among the family of these discoid remains a "*Beltanelliformis*-type", characterized by the presence of concentric peripheral fine wrinkles, and a "*Nemiana*-type", more globose and smooth, corresponding to two taphonomic processes of the same original organism of undefined nature (Narbonne, 2007). On the other hand, Leonov (2007a) attributes these two types of preservation to two different organisms: a "*Beltanelloides*" form that would be attributable to a planktonic spherical organism and a "*Nemiana*" form that would result from the dwelling imprint of a benthic bag-shaped organism. It should be noted that this distinction is based in part on measurements showing a significant difference between the two "morphotypes". The differences in size between our two specimens show that it cannot be a discriminating argument (Fig. 4). The dilemma has been convincingly summarized by the recent comprehensive revision of Ivantsov et al. (2014), which demonstrates that, due to the synonymy and anteriority aspects, our fossils must be better related to *Beltanelliformis brunsae* Menner in Keller, 1974. The name

*Beltanelloides sorichevae* is thus not valid, as indicated by Narbonne & Hofmann (1987) and other authors having often pointed out in synonymy the two species (Fedonkin & Runnegar 1992). As a result, the studied specimens are determined as follow (for complete diagnosis and synonymy, refer to Ivantsov *et al*. 2014):

Regnum incertae

Genus *Beltanelliformis* Menner in Keller et al., 1974

*Beltanelliformis brunsae* Menner in Keller et al., 1974

It should be noted that in this work *Beltanelloides sorichevae* Sokolov, 1965 and *Nemiana simplex* Palij, 1976 are clearly synonymized with *Beltanelliformis brunsae*.

**DISCUSSION**

The problem of systematic assignment is closely linked, not only to the taphonomic processes themselves, but also to the inferred original organic nature of these fossils. Thus, the remains comparable to our specimens could be considered at the same time to be fossil bodies, fossil traces or megascopic compression (Hofmann 1992a, 1992b; Runnegar & Fedonkin 1992, Fedonkin & Runnegar 1992, Runnegar 1992a, 1992b; Jensen *et al*. 2006). Like many other discoidal elements of the Ediacaran biota, *Beltanelliformis* was first considered as "medusoid" organism (Sokolov 1972; Palij *et al*. 1979; Fedonkin 1981; Sokolov & Fedonkin 1984). As mentioned above, "*Nemiana*-type" preservation has been attributed to bug-shaped organisms. For a long time an affinity with benthic coelenterates such as anemones was considered. Various reconstructions have favored this option, these fossils being clearly ranked among the coelenterates (Gureev 1985; Fedonkin 1990; Fedonkin 1992; Fedonkin 1994; Seilacher *et al.* 2005). On other hand, according to Leonov (2007a), the "*Beltanelloides*-type" remains would correspond to spherical floating organisms, formed of a thin and flexible envelope fallen to the bottom and whose compaction would explain the fine concentric lines around the periphery. In the same sense, Ivantsvov et al. (2014) consider that the presence of basically plastic prediagenetic distortions, as observed for Romanian specimens, suggests that the envelope of *Beltanelloides* was also elastic. If we adopt the idea of a single type of organism, we must reconcile the different types of preservation. Narbonne & Hofmann (1987) proposed a scenario arguing a continuum, but they give no conclusion as to the exact nature of the organism.

Flat discoid fossils, like that of Dobrogea, have often been interpreted as compressions of more or less spherical organisms, of which, in some cases, there are only traces or organic films, or more rarely both. As a result, they have been compared with other *Chuaria*-type Ediacaran fossils known as carbon compressions, algal or microbial in origin (Hofmann 1994; Steiner 1996; Steiner & Reitner 2001; Leonov 2004; Ragozina & Leonov 2004; Grazhdankin *et al.* 2005; Xiao & Dong 2006; Grazhdankin *et al*. 2007; Leonov 2007a, 2007b; Leonov & Ragozina 2007; Moczydłowska 2008; Ragozina *et al*. 2016; Bykova 2017; Ye *et al.* 2017), although this affinity is doubtful for other authors (Narbonne & Hofmann 1987). According to Xiao *et al.* (2017), stable carbon isotope values for samples of *Beltanelliformis* preserved as carbonaceous macrofossil do not allow to discriminate between interpretations of this organism as a colonial bacterium or a eukaryotic alga. However, the most recent studies dedicated to biomarkers from *Beltanelliformis* specimens similar to those of Romania, but with a preserved organic film, favor a microbial origin, probably cyanobacterial (Bobrovsky *et al.* 2016, Bobrovsky *et al*. 2017, Ivantsov 2017).

The known occurrences of *Beltanelliformis brunsae* are apparently limited exclusively to the Ediacaran period (=Vendian), being perhaps the most common and widely distributed Ediacaran fossil worldwide (Narbonne 1998; McCall 2006). In the Precambrian of Russia (Siberia, Urals, White Sea ...) and Ukraine, this common species seems characteristic of the "Upper Vendian" (see review in Ivantsov *et al*. 2014). In other parts of the world, known occurrences also correspond to the upper Neoproterozoic: Great Britain (Pyle *et al.* 2004; McIllroy *et al*. 2005; Liu 2011), Canada and Newfoundland (Hofmann 1983; Liu *et al.* 2015b), China (Zhao *et al.* 2004; Wang *et al.* 2014), Mongolia (Ragozina *et al.* 2016), South America (Aceñolaza & Alonso 2001; Drefahl & Silva 2007; Netto 2012), Namibia (Wood *et al*. 2015), Australia (Xiao *et al.* 2013). Consequently, Romanian specimens extend westward the distribution field of the species known in East Europa. They also testify to an Ediacarian age for the Histria formation of central Dobrogea.

Occurrences of *Beltanelliformis* around the world generally concern sediments deposited in shallow water environment (Narbonne & Hofmann, 1987; Fedonkin, 1992; Narbonne, 1998; Aceñolaza & Alonso, 2001; Grazhdankin 2004; Pyle *et al*. 2004; Grazhdankin *et al*. 2005; McIllroy *et al*. 2005; Seilacher *et al*. 2005; Leonov 2007a; Narbonne 2007; Rozhnov 2009; Rozhnov 2010; Liu,2011; Ivantsov *et al*. 2014; Ivantsov 2017; Liu *et al.* 2015b; Netto 2012; Grytsenko 2016; Ragozina *et al*. 2016), sometimes in connection with storm deposits and hummocky cross-stratification (HCS) (Narbonne & Hofmann, 1987; Pyle et al., 2004; McCall, 2006; Narbonne, 2007). If the cyanobacterial nature is proved, we can estimate that this would correspond to the photic zone. This trend is consistent with the microbial mat surfaces observed in the Histria Formation (Saint Martin *et al*. 2011; Saint Martin *et a*l. 2012) and the sedimentological features of the specimen 2.

## CONCLUSIONS

The studied samples unambiguously present exactly the same characteristics as various samples described in the Ediacarian deposits in Podolia, White Sea or Siberia. Although the discussion of their origin is not yet fully decided, recent studies seem to show that the hypothesis of the cyanobacterial colony imprint can be favored. The attribution of these specimens to the Ediacaran morpho-species *Beltanelliformis brunsae* Menner in Keller eta al., 1974 is probably the most acceptable.

The most interesting consequence is the confirmation of the Ediacaran age of the Histria Formation and therefore of the various organic traces known in Central Dobrogea. The deposits of the Histria Formation therefore represent a certain potential for restitution of the Ediacarian life in the Romanian ground. Further systematic field investigations would better measure the abundance and the diversity of the Ediacaran fossils.

## ACKNOWLEDGEMENTS

We address special thanks to Gheorghe Oaie, dear to our memory, who opened the way for the work on the organic remains of Neoproterozoic from Dobrogea and has always supported our investigations. We are also indebted to Antoneta Seghedi who shared with us his perfect knowledge of the field and facilitated the study of the specimen of the Museum of Geology. This work was carried out thanks to the French-Romanian

**REFERENCES**

ACEÑOLAZA F.G. & ALONSO Y.R.N. 2001. – Icno-asociaciones de la transición Precámbrico /Cámbrico en el noroeste de Argentina. *Journal of Iberian Geology* 27: 11-22.

ANTCLIFFE J. & BRASIER M. 2008. – *Charnia* at 50: developmental models for Ediacaran fronds. *Palaeontology* 51 (1): 11-26.

ANTCLIFFE J.B. & HANCY A.D. 2013. – Critical questions about early character acquisition-Comment on Retallack 2012: "Some Ediacaran fossils lived on land". *Evolution* & *Development* 15 (4): 225-227.

ANTCLIFFE J.B., HANCY A.D. & BRASIER M.D. 2015. – A new ecological model for the ∼ 565 Ma Ediacaran biota of Mistaken Point, Newfoundland. *Precambrian Research* 268: 227-242.

AROURI K., CONAGHAN P.J., WALTER M.R., BISCHOFF G.C.O. & GREY K. 2000. – Reconnaissance sedimentology and hydrocarbon biomarkers of Ediacarian microbial mats and acritarchs, lower Ungoolya Group, Officer Basin. *Precambrian Research* 100: 235-280.

BALINTONI I., BALICA C., SEGHEDI A. & DUCEA M. 2011. – Peri-Amazonian provenance of the Central Dobrogea terrane (Romania) attested by U/Pb detrital zircon age patterns. *Geologica Carpathica* 62 (4): 299-307.

BOBROVSKIY I., HOPE J.M., LIYANAGE T.M. & BROCKS J.J. 2016. – Biomarkers from the Ediacaran macrofossil *Beltanelliformis*. 9[th] Australian Organic Geochemistry Conference, Fremandle, Australia, Abstracts, pp. 50-51.

BOBROVSKIY I., HOPE J.M. & BROCKS J.J. 2017. – Biomarkers analysis of the Ediacaran macrofossil *Beltanelliformis*. Goldschmidt Conference , Paris 2017, Abstracts, https://goldschmidt.info/2017/abstracts/abstractView?id=2017003569.

BOUOUGRI E.H. & PORADA H. 2007. – Siliciclastic biolaminites indicative of widespread microbial mats in the Neoproterozoic Nama Group of Namibia. *Journal of African Earth Sciences* 48: 38-48.

BOWYER F., WOOD R.A. & POULTON S.W. 2017. – Controls on the evolution of Ediacaran metazoan ecosystems: A redox perspective. *Geobiology* 15: 516-551.

BRASIER M. & ANTCLIFFE J., 2004. – Decoding the Ediacaran enigma. *Science* 305: 115-116.

BRASIER M. & ANTCLIFFE J. 2009. – Evolutionary relationships within the Avalonian Ediacara biota: new insights from laser analysis. *Journal of the Geological Society, London* 166: 363-384.

BRASIER M.D., ANTCLIFFE J.B. & LIU A.G., 2012. – The architecture of Ediacaran fronds. Palaeontology, 55(5): 1105-1124.

BRIGGS D.E.G., 2017. – Seilacher, Konstruktions-Morphologie, morphodynamics, and the evolution of form. *Journal of Experimental Zoology Part B: Molecular and Developmental Evolution* 328B: 197-206.

BUDD G.E. & JENSEN S. 2017. – The origin of the animals and a 'Savannah' hypothesis for early bilaterian evolution. *Biological Review* 92: 446-473.

BYKOVA N. 2017. – Paleoecology and Taphonomy of Ediacaran macrofossils from the Khatyspyt Formation, Olenek Uplift, Siberia. PhD, Virginia Polytechnic Institute and State University, 125 p.

BYKOVA N., GILL B.C., GRAZHDANKIN D., ROGOV V. & XIAO S. 2017. – A geochemical study of the Ediacaran discoidal fossil *Aspidella* preserved in limestones: Implications for its taphonomy and paleoecology. *Geobiology* 15: 572-587.

CALLOW R.H.T. & BRASIER M. 2009. – Remarkable preservation of microbial mats in Neoproterozoic siliciclastic settings: Implications for Ediacaran taphonomic models. *Earth-Science Reviews* 96: 207-219.

CUI H., GRAZHDANKIN D.V., XIAO S., PEEK S., ROGOV V., BYKOVA N.V., SIEVERS N.E., LIU X.M. & KAUFMAN A.J. 2016. – Redox-dependent distribution of early macro-organisms: Evidence from the terminal Ediacaran Khatyspyt Formation in Arctic Siberia. *Palaeogeography, Palaeoclimatology, Palaeoecology* 461: 122-139.

DECECCHI T.A., NARBONNE G.M., GREENTREE C. & LAFLAMME M. 2017. – Relating Ediacaran fronds. *Paleobiology* 43(2): 171-180.

DREFAHL M. & SILVA R. C. 2007. – Ocorrências de fósseis ediacaranos na Formação Camarinha (Neoproterozóico Superior), Sudeste do Estado do Paraná. In: ISMAR de SOUZA CARVALHO et al. (eds). Paleontologia: Cenários da Vida. Rio de Janeiro: Editora Interciência Ltda., v. 1, pp. 3-11.

DROSER M.L. & GEHLING J.G. 2015. – The advent of animals: The view from the Ediacaran. *Proceedings of the National Academy of Sciences* 112(16): 4865-4870.

DUFOUR S.C. & McILROY D. 2016. – Ediacaran pre-placozoan diploblasts in the Avalonian biota: the role of chemosynthesis in the evolution of early animal life. In: BRASIER A. T., MCILROY D. & MCLOUGHLIN N. (eds) Earth System Evolution and Early Life: a Celebration of the Work of Martin Brasier. Geological Society, London, Special Publications, 448 (1), pp. 211-219.

ELLIOTT D.A., VICKERS-RICH P., TRUSLER P. & HALL M. 2011. – New evidence on the taphonomic context of the Ediacaran *Pteridinium*. *Acta Palaeontologica Polonica* 56(3): 641-650.

ERWIN D. H. & TWEEDT S. 2011. – Ecological drivers of the Ediacaran-Cambrian diversification of Metazoa. *Evolutionary Ecology* 26(2): 417-433.

FEDONKIN M.A. 1981. – The Vendian White Sea biota (Precambrian skeletonless fauna of the northern Russian Platform), *Trudy* Geologiceskogo Instituta, Akademija *Nauk* SSSR 342: 1-99.

FEDONKIN M.A. 1990. – Systematic description of Vendian metazoa. In: SOKOLOV B.S. & IWANOWSKI A.B. (eds) The Vendian system. Vol. 1 Paleontology, Springer-Verlag, pp. 71-120.

FEDONKIN M.A. 1992. – Vendian faunas and the early evolution of Metazoa, In: LIPPS J.H. & SIGNOR P.W. (Eds) Origin and the Early Evolution of Metazoa, New York, Plenum Press, pp. 87-129.

FEDONKIN M.A. 1994. – Vendian body fossils and trace fossils. In: BENGTSON. S. (ed.) Early Life on Earth. Nobel Symposium No. 84. Columbia U.P., New York, pp. 370-388.

FEDONKIN M.A. & RUNNEGAR B.N., 1992. – Proterozoic metazoan trace fossils. In: SCHOPF J.W. & KLEIN C. (eds) The Proterozoic Biosphere-a Multidisciplinary Study, New York, Cambridge Univ. Press, pp. 389-395.

FEDONKIN M.A. & VICKERS-RICH P. 2007. – The White Sea's Windswept Coasts. In: FEDONKIN M.A., GEHLING J.G., GREY K., NARBONNE G.M. & VICKERS-RICH P. (eds) The Rise of Animals, Johns Hopkins University Press, Baltimore, pp. 115-156.

FEDONKIN M.A., GEHLING J.G., GREY K., NARBONNE,G.M. & VICKERS-RICH P. 2007. – The Rise of Animals, Johns Hopkins University Press, Baltimore, 326 pp.

GLAESSNER M.F. 1984. – The dawn of animal life. A biohistorical study. Cambridge University Press, 182 p.

GEHLING J.G. & DROSER M.L. 2009. – Textured organic surfaces associated with the Ediacara biota in South Australia. *Earth-Science Reviews* 96, 196-206.

GIUŞCĂ D., IANOVICI V., MÎNZATU S., SOROIU E., LEMNE M., TĂNĂSESCU A. & IONCICĂ M. 1967. – Asupra virstei absolute a formatiunilor cristaline din vorlandul orogenului carpatic. A*cademia Republicii Socialiste România, Studii si Cercetari de Geologie, Geofizica, Geografie, seria Geologie* 12(2): 287-296.

GRAZHDANKIN D. 2004. – Patterns of distribution in the Ediacaran biotas: Facies versus biogeography and evolution. *Paleobiology* 30: 203-221.

GRAZHDANKIN D. V. 2014. – Patterns of evolution of the Ediacaran soft-bodied biota. *Journal of Paleontology* 88 (2): 269-283.

GRAZHDANKIN D. V. & GERDES G. 2007. – Ediacaran microbial colonies. *Lethaia* 40: 201-210.

GRAZHDANKIN D.V., MASLOV A.V., MUSTILL T.M.R. & KRUPENIN M.T. 2005. – The Ediacaran White Sea biota in the Central Urals. *Doklady Earth Sciences* 401A (3): 382-385.

GRAZHDANKIN D.V., NAGOVITSIN K.E. & MASLOV A.V. 2007. – Late Vendian Miaohe-type ecological assemblage of the East European Platform. *Doklady Earth Sciences* 417 (8): 1183-1187.

GRYTSENKO V. 2014. – A new discovery of metazoa imprints and ichnofossils in the Vendian Mohyliv suite from the Bernashivka quarry. *Proceedings of the National Museum of Natural History, Ukraine* 14: 23-34.

GUREEV Y.A. 1985. – Vendiata-primitive Precambrian Radialia. In: SOKOLOV B.S. & ZHURAVLEVA I.T. (eds) Problematics of the Late Precambrian and Paleozoic. Nauka, Moscow, pp. 92-103.

HOFMANN H.J. 1992a. – Proterozoic carbonaceous films. In: SCHOPF J.W. & KLEIN C. (eds) The Proterozoic Biosphere-a Multidisciplinary Study, New York, Cambridge Univ. Press, pp. 349-357.

HOFMANN H.J. 1992b. – Proterozoic and selected Cambrian megascopic carbonaceous films. In: SCHOPF J.W. & KLEIN C. (eds) The Proterozoic Biosphere-a Multidisciplinary Study, New York, Cambridge Univ. Press, pp. 957-979

HOFMANN H.J. 1994. – Proterozoic carbonaceous compressions ("metaphytes" and "worms"). In: BENGTSON S. (ed.) Early life on Earth. Nobel Symposium 84, Columbia U.P., New York, pp. 342-357.

HOFMANN H. J., FRITZ W. H. & NARBONNE G.M. 1983. – Ediacaran (Precambrian) fossils from the Wernecke Mountains, Northwestern Canada, *Science* 221 (4609): 455-457.

IVANTSOV A.Y. 2017. – The most probable Eumetazoa among late Precambrian macrofossils. *Invertebrate Zoology* 14 (2): 127-133.

IVANTSOV A.Y. 2017. – Finds of Ediacaran-Type fossils in Vendian deposits of the Yudoma Group, Eastern Siberia. Doklady Earth Sciences 472(2): 143-146.

IVANTSOV A.Y., GRITSENKO V. P., KONSTANTINENKO L.I. & ZAKREVSKAYA M.A. 2014. – Revision of the Problematic Vendian Macrofossil *Beltanelliformis* (=*Beltanelloides*, *Nemiana*). *Paleontological Journal* 48 (13): 1-26.

JENSEN S., DROSER M.L. & GEHLING J.G. 2006. – A critical look at the Ediacaran trace fossil record. In: XIAO S. & KAUFMAN A.J. (eds). Neoproterozoic Geobiology and Paleobiology, Springer, 115-157.

KELLER B.M., MENNER V.V., STEPANOV V.A., & CHUMAKOV N.M. 1974. – New finds of fossils in the Precambrian Valdai Series along the Syuzma River. *Izvestia Akademiya Nauk SSSR, Seriya Geologiya* 12: 130-134.

KENCHINGTON C.G. & WILBY P.R. 2014. – Of time and taphonomy: preservation in the Ediacaran. In: LAFLAMME M., SCHIFFBAUER J.D. & DARROCH S.A.F. (eds.) Reading and writing of the fossil record: preservational pathways to exceptional fossilization. The Paleontological Society Papers 20, The Paleontological Society Short Course, pp. 101-122.

KOLESNIKOV A.V., DANELIAN T., GOMMEAUX M., MASLOV A.V. & GRAZHDANKIN D.V. 2017. – Arumberiamorph structure in modern microbial

mats: implications for Ediacaran palaeobiology. *Bulletin de la Société géologique de France* 188(1-2): 57-66.

KRÄUTNER H.G., MUREȘAN M. & SEGHEDI A. 1988. – Precambrian of Dobrogea. In: ZOUBEK V. (ed), Precambrian in Younger Fold Belts, John Wiley, New York, pp. 361-379.

KUMAR S. & AHMAD S. 2014. – Microbially induced sedimentary structures (MISS) from the Ediacaran Jodhpur Sandstone, Marwar Supergroup, western Rajasthan. *Journal of Asian Earth Sciences* 91: 352-361.

LAFLAMME M. & NARBONNE G. 2008. – Ediacaran fronds. *Palaeogeography, Palaeoclimatology, Palaeoecology* 258: 162-179.

LAFLAMME M., SCHIFFBAUER J.D., NARBONNE G.M. & BRIGGS D.E.G. 2010. – Microbial biofilms and the preservation of the Ediacara biota. *Lethaia* 44 (2): 203-213.

LAFLAMME M., XIAO S. & KOWALEWSKI M. 2009. – Osmotrophy in modular Ediacara organisms. *Proceedings of the National Academy of Sciences* 106: 14438-14443.

LAN Z.W. & CHEN Z.Q. 2013. – Proliferation of MISS-forming microbial mats after the late Neoproterozoic glaciations: Evidence from the Kimberley region, NW Australia. *Precambrian Research* 224: 529-550

LEONOV M.V. 2004. – Comparative taphonomy of the Vendian genera *Beltanelloides* and *Nemiana* as a key to their true nature. Abstract volume for IGCP493 Workshop – The Rise and Fall of the Vendian Biota, Prato, pp. 66-71, + poster.

LEONOV M.V. 2007a. – Comparative taphonomy of Vendian genera *Beltanelloides* and *Nemiana*: taxonomy and lifestyle. In: VICKERS-RICH P & KOMAROWER P.

(eds) The Rise and fall of the Ediacaran Biota. Geological Society, London, Special Publications, 286, pp. 259-267.

LEONOV M. 2007b. – Macroscopic plant remains from the base of the Ust'-Pinega Formation (Upper Vendian of the Arkhangelsk Region). *Paleontological Journal* 41 (6): 683-691.

LEONOV M.V. & RUD'KO S.V., 2012. – A find of Vendian fossils in the beds of the Dal'nyaya Taiga Group of the Patoman Plateau. *Stratigraphy and Geological Correlation* 20 (5): 96-99.

LEONOV M.V. & RAGOZINA A.L. 2007. – Upper Vendian assemblages of carbonaceous micro- and macrofossils in the White Sea Region: systematic and biostratigraphic aspects. In: VICKERS-RICH P. & KOMAROWER P. (eds) The Rise and Fall of the Ediacaran Biota, Geological Society, London, Special Publications, 286, pp. 269-275.

LIU A.G. 2011. – Reviewing the Ediacaran fossils of the Long Mynd, Shropshire. *Proceedings of the Shropshire Geological Society* 16: 31-43.

LIU A.G., MCILROY D. & BRASIER M.D. 2010. – First evidence for locomotion in the Ediacara biota from the 565 Ma Mistaken Point Formation, Newfoundland. *Geology* 38 (2): 123-126.

LIU A.G., KENCHINGTON C.G. & MITCHELL,E.G. 2015a. – Remarkable insights into the paleoecology of the Avalonian Ediacaran macrobiota. *Gondwana Research* 27: 1355-1380.

LIU A.G., MATTHEWS J.J., HERRINGSHAW L.G. & MCILROY D. 2015b. – Mistaken Point Ecological Reserve Field Trip Guide. In McILROY D. (ed.):

Ichnology: Papers from Ichnia III, Edition: Miscellaneous Publication 9, Chapter: Part B, Geological Association of Canada, pp. 231-272.

McCALL G. 2006. – The Vendian (Ediacaran) in the geological record: Enigmas in geology's prelude to the Cambrian explosion. *Earth-Science Reviews* 77: 1-229.

McGABHANN B.A. 2014. – There is no such thing as the 'Ediacara Biota'. *Geoscience Frontiers* 5: 53-62.

McILROY D., BRASIER D. & LANG A., 2009. – Smothering of microbial mats by macrobiota: implications for the Ediacara biota. *Journal of the Geological Society, London* 166: 1117-1121.

McILROY D., CRIMES T.P. & PAULEY J.C. 2005. – Fossils and matgrounds from the Neoproterozoic Longmyndian Supergroup, Shropshire, *U.K. Geological Magazine* 142: 441-455.

McMENAMIN M.A.S. 1986. – The Garden of Ediacara. *Palaios* 1 (2): 178-182.

MOCZYDŁOWSKA M. 2008. – The Ediacaran microbiota and the survival of Snowball Earth conditions. *Precambrian Research* 167: 1-15.

NARBONNE G.M. 1998. – The Ediacara Biota: A Terminal Neoproterozoic experiment in the evolution of life. *GSA Today* 8 (2): 1-6.

NARBONNE G.M., 2007. – The Canadian Cordillera. In: FEDONKIN M.A., GEHLING J.G., GREY K., NARBONNE G.M. & VICKERS-RICH P. (eds) The Rise of Animals, Johns Hopkins University Press, Baltimore, pp.

NARBONNE G.M. & HOFMANN H.J., 1987. – Ediacaran biota of the Wernecke Mountains, Yukon, Canada. *Palaeontology* 30 (4): 647-676.

NARBONNE G.M., LAFLAMME M., TRUSLER P.W., DALRYMPLE R.W. & GREENTREE C., 2014. – Deep-water Ediacaran fossils from Northwestern Canada: taphonomy, ecology, and evolution. *Journal of Paleontology* 88 (2): 207-223.

NETTO RG. 2012. – Evidences of life in terminal Proterozoic deposits of southern: a synthesis. In: NETTO RG, CARMONA NB & TOGNOLI FMW (Eds). Ichnology of Latin America - selected papers. Porto Alegre: SBP, Monografias da Sociedade Brasileira de Paleontologia, pp. 15-26.

OAIE G. 1992. – Traces of organic activity in the Greenschist Series of central Dobrogea (Romania). *Studii şi Cercetări de Geologie* 37: 77-81.

OAIE G. 1993. – Necesitatea protejării urmelor de actitate organică de vârstă precambrian superior din cadrul seriei şisturilor verzi din Dobrogea centrală. *Ocrotirea Naturiişi a Mediului în conjurator* 37 (2): 133-137.

OAIE G. 1998. – Sedimentological significance of mudstone microclast intervals in Upper Proterozoic turbidites, Central Dobrogea, Romania. *Sedimentary Geology* 115: 289-300.

OAIE G. 1999. – Sedimentologia şi tectonica seriei Şisturilor Verzi din Dobrogea Centrală şi prelungireaei în acvatoriul MăriiNegre. Teza de doctorat, Universitatea din Bucureşti, 105 p.

OAIE G., SEGHEDI A., RĂDAN S. & VAIDA M., 2005. – Sedimentology and source area composition for the Neoproterozoic-Eocambrian turbidites from East Moesia. *Geologica Belgica* 8 (4): 78-105.

OAIE G., SEGHEDI A., SAINT MARTIN J.P. & SAINT MARTIN S., 2012. – Signification sédimentologique des traces mécaniques et biogènes des dépôts précambriens et paléozoïques de Dobrogea centrale et du nord. *GeoEcoMarina, supplement* 18, 38-39.

PALIJ V.M. 1976. – Remains of a skeletonless fauna and traces of life ability from deposits of the Upper Precambrian and Lower Cambrian of Podolia. In: RYABENKO V.A. (ed.) Paleontology and stratigraphy of the Upper Precambrian and Lower Paleozoic of the southwestern East European Platform, Kiev, Naukova Dumka, pp. 63-76.

PALIJ V.M., POSTI E. & FEDONKIN M.A., 1979. – Soft body Metazoa and fossil animal's traces of the Vendian and Early Cambrian. In: KELLER B.M. & ROZANOV,A.Y. (eds) Paleontology of the Upper Precambrian and Cambrian Deposits of the East European Platform, Moscow, Nauka, pp. 49-82.

PARIHAR V.S., GAUR V., NAMA S.L. & MATHUR S.C., 2015. – New report of *Arumberia banksi* (Ediacaran affinity) Mat Structures from the Girbhakar Sandstone of Marwar Supergroup, Bhopalgarh area, Jodhpur, Western Rajasthan, India. In: SHRIVASTAVA, K.L. & SRIVASTAVA, P.K. (eds.) Frontiers of Earth Science, Scientific Publishers (India), pp. 385-392.

PETERSON K.J., COTTON J.A., GEHLING J.G. & PISANI D., 2008. – The Ediacaran emergence of bilaterians: congruence between the genetic and the geological fossil records. *Philosophical Transactions of the Royal Society* B 36: 1435-1443.

PYLE L.J., NARBONNE G.M., JAMES N.P., DALRYMPLE R.W. & KAUFMAN, A.J. 2004. – Integrated Ediacaran chronostratigraphy, Wernecke Mountains, northwestern Canada. *Precambrian Research* 132: 1-27.

RAGOZINA A. L. & LEONOV M.V., 2004. – Macrophytes and organically preserved microfossils in the Vendian complex of the southeastern White Sea area, Russia.

Abstract volume for IGCP493 Workshop - The Rise and fall of the Vendian Biota, Prato, pp. 86-87, + poster.

RAGOZINA A.L., DORJNAMJAA D., SEREZHNIKOVA E.A., ZAITSEVAA L.V. & ENKHBAATAR B. 2016. – Association of macro- and microfossils in the Vendian (Ediacaran) postglacial successions in Western Mongolia. *Stratigraphy and Geological Correlation* 24(3): 242-251.

REID L.M., GARCÍA-BELLIDO D.C., PAYNE J.L., RUNNEGAR B. & GEHLING J.G. 2017. – Possible evidence of primary succession in a juvenile-dominated Ediacara fossil surface from the Flinders Ranges, South Australia. *Palaeogeography, Palaeoclimatology, Palaeoecology* 476: 68-76.

RETALLACK G.J. 1994. – Were the Ediacaran fossils lichens? Paleobiology, 20: 523-544.

RETALLACK G.J. 2010. – First evidence for locomotion in the Ediacara biota from the 565 Ma Mistaken Point Formation, Newfoundland: Comment. *Geology* doi:10.1130/G31137C.1.

RETALLACK G.J. 2013a. – Ediacaran characters. *Evolution & Development* 15 (6): 387-388.

RETALLACK G. J. 2013b. – Ediacaran life on land. *Nature* 493: 89-92.

RETALLACK G. J. 2016. – Ediacaran sedimentology and paleoecology of Newfoundland reconsidered. *Sedimentary Geology* 333: 15-31.

ROZHNOV S.V. 2009. – Development of the trophic structure of Vendian and Early Paleozoic marine communities. *Paleontological Journal* 43 (11): 1364-1377.

ROZHNOV S.V. 2010. – From Vendian to Cambrian: the beginning of morphological disparity of modern metazoan phyla. *Russian Journal of Developmental Biology* 41(6): 357-368.

RUNNEGAR B.N. 1992a. – Proterozoic fossils of soft-body metazoan (Ediacara faunas). In: SCHOPF, J.W. & KLEIN, C. (eds) The Proterozoic Biosphere-a Multidisciplinary Study, New York, Cambridge Univ. Press, pp. 999-1007.

RUNNEGAR B.N. 1992b. – Proterozoic metazoan trace fossils. In: SCHOPF J.W. & KLEIN C. (eds) The Proterozoic Biosphere-a Multidisciplinary Study, New York, Cambridge Univ. Press, pp. 1009-1015.

RUNNEGAR B.N. & FEDONKIN M.A. 1992. – Proterozoic metazoan body fossils. In: SCHOPF J.W. & KLEIN C. (eds) The Proterozoic Biosphere-a Multidisciplinary Study, New York, Cambridge Univ. Press, pp. 369-388.

SAINT MARTIN,J.P., CHARBONNIER S., SAINT MARTIN S., OAIE G., SEGHEDI A. & RICHIR P., 2011. – Evidence of microbial mats in the Ediacaran deposits of Dobrogea (Romania). *GeoEcoMarina, Supplement* 17, 151-152.

SAINT MARTIN J.P., FERNANDEZ S., OAIE G., SEGHEDI A., SAINT MARTIN S., CHARBONNIER S. & ANDRE J.P. 2013 – Le monde ediacarien de Dobrogea. In : SAINT MARTIN J.P. (ed.) Recherches croisées en Dobrogea, Editura Amanda Edit, Bucarest, 29-39.

SAINT MARTIN J.P., SAINT MARTIN S., OAIE G. & SEGHEDI A. 2012. – Traces of organic activity and microbial mats in the Ediacaran basement of the Moesian Platform, Romania. 34th International Geological Congress, Brisbane, Australia, Abstracts, 1 p.

SCHIFFBAUER J. D. BYKOVA N., &. MUSCENTE A.D. 2017. – Fossils, proxies, and models: Geobiology at critical transitions in the Proterozoic–Paleozoic. *Geobiology* 15 (4): 467-468.

SEGHEDI A. 2012. – Palaeozoic Formations from Dobrogea and Pre-Dobrogea - An Overview. *Turkish Journal of Earth Sciences* 21: 669-721.

SEGHEDI A. & OAIE G., 1995. – Palaeozoic evolution of North Dobrogea. In: SĂNDULESCU M., SEGHEDI A., OAIE G., GRĂDINARU E. & RĂDAN S. (eds.) Field Guidebook, Central and North Dobrogea. IGCPProject No. 369 "Comparative evolution of Peri-Tethyan Rift Basins", Mamaia 1995, pp. 1-75.

SEGHEDI A., BERZA T., IANCU V., MĂRUNTIU M. & OAIE G. 2005. – Neoproterozoic terranes in the Moesian basement and in the Alpine Danubian nappes of the South Carpathians. *Geologica Belgica* 8 (4): 4-19.

SEILACHER A. 1989. – Vendozoa: organismic construction in the Proterozoic biosphere. *Lethaia* 22: 229-239.

SEILACHER A. 1992. – Vendobionta and Psammocorallia: lost constructions of Precambrian evolution. *Journal of the Geological Society, London* 149 (4): 607-613

SEILACHER A., BUATOIS L.A. & MÁNGANO M.G. 2005. – Trace fossils in the Ediacaran-Cambrian transition: Behavioral diversification, ecological turnover and environmental shift. *Palaeogeography, Palaeoclimatology, Palaeoecology* 227: 323-356.

SEREZHNIKOVA E.A. 2010. – Colonization of substrates: Vendian sedentary benthos. *Paleontological Journal 44* (12): 1560-1569.

SHEN B., DONG L., XIAO S. & KOWALEWSKI M. 2008. – The Avalon explosion: evolution of Ediacara morphospace, *Science* 319 (5859): 81-84.

SHU, D., 2008. – Cambrian explosion: Birth of tree of animals. *Gondwana Research* 14: 219-240.

SOKOLOV B.S. 1965. – Most ancient deposits of the Early Cambrian and sabelliditids. In: All-Union Symposium on Paleontology of the Precambrian and Early Cambrian, Novosibirsk, Inst. Geol. Geofiz. Sib. Otd. Akad. Nauk SSSR, pp. 78-91.

SOKOLOV B.S. 1972. – The Vendian Stage in Earth History, in: Twenty-Fourth Session of the International Geological Congress. Section 1. Precambrian Geology. Montréal, pp. 78-84.

SOKOLOV B.S. 1976. – Organic world of the Earth on the way to Phanerozoic differentiation, *Vest. Akademika Nauk SSSR,* 1: 126-143.

SOKOLOV B.S. 1997. – Essays on the Advent of the Vendian System. Moscow: KMK Scientific Press Ltd, 142 pp.

SOKOLOV B.S. & FEDONKIN M.A. 1984. – The Vendian as the terminal system of the Precambrian. *Episodes* 7 (1): 12-19.

SOKOLOV B.S. & IWANOWSKI A.B. 1990. – The Vendian system. Vol. 1 Paleontology, Springer-Verlag, 383 pp.

STEINER M., 1996. – *Chuaria circularis* Walcott 1899-''megasphaeromorph acritarch'' or prokaryotic colony? *Acta Universitatis Carolinae Geologica* 40: 645-665.

STEINER M. & REITNER J. 2001. – Evidence of organic structures in Ediacara-type fossils and associated microbial mats. *Geology* 29 (12): 1119-1122.

TARHAN L.G. 2010. – New morphological diversity or preservational variability?: resolving the taphonomic context of Ediacaran assemblage-scale heterogeneity. *Palaios* 25: 823-830.

TARHAN L.G. & LAFLAMME M. 2015. – An examination of the evolution of Ediacaran paleoenvironmental and paleoecological research. *Palaeogeography, Palaeoclimatology, Palaeoecology* 434:1-3

TARHAN L.G., DROSER M.L. & GEHLING J.G. 2010. – Taphonomic controls on Ediacaran diversity: uncovering the holdfast origin of morphologically variable enigmatic structures. *Palaios* 25: 823-830.

TARHAN L.G., DROSER M., GEHLING J.G. & DZAUGIS M.P. 2017. – Microbial mat sandwiches and other anactualistic sedimentary features of the Ediacara member (Rawnsley quartzite, South Australia): implications for interpretation of the Ediacaran sedimentary record. *Palaios* 32: 181-194.

WANG Y., WANG Y, DU W & WANG X., 2014. – The correlation between macroscopic algae and metazoans in the Ediacaran: a case study on the Wenghui biota in northeastern Guizhou, South China. *Australian Journal of Earth Sciences* 61: 967-977.

WOOD R.A., POULTON S.W., PRAVE A.R., HOFFMANN K.H.; CLARKSON M.O., GUILBAUD R., LYNE J.W., TOSTEVIN R., BOWYER F., PENNY A.M., CURTIS A. & KASEMAN S.A., 2015. – Dynamic redox conditions control late Ediacaran metazoan ecosystems in the Nama Group, Namibia. *Precambrian Research* 261: 252-271.

XIAO S. & DONG L. 2006. – On the morphological and ecological history of Proterozoic Macroalgae. In: XIAO, S. & KAUFMAN, A.J. (eds.) Neoproterozoic Geobiology and Paleobiology, Springer, pp. 57-90.

XIAO S. & LAFLAMME M. 2008. – On the eve of animal radiation: phylogeny, ecology and evolution of the Ediacara biota. *Trends in Ecology and Evolution* 24: 31-40.

XIAO S., BYKOVA N., KOVALICK A. & GILL B.C. 2017. – Stable carbon isotopes of sedimentary kerogens and carbonaceous macrofossils from the Ediacaran Miaohe Member in South China: Implications for stratigraphic correlation and sources of sedimentary organic carbon. *Precambrian Research* 302: 171-179.

XIAO S., DROSER M., GEHLING J.G., HUGHES I.V., WAN B., CHEN Z. & YUAN X. 2013. – Affirming life aquatic for the Ediacara biota in China and Australia, *Geology* 41 (10): 1095-1098.

XIAO S., YUAN X., STEINER M. & KNOLL A.H. 2002. – Macroscopic carbonaceous compressions in terminal Proterozoic shale: A systematic reassessment of the Miaohe biota, South China. *Journal of Paleontology* 76: 247-376.

YE Q., TONGA J., ANC Z., HUA J., TIANA L., GUANA K. & XIAO S. 2017. – A systematic description of new macrofossil material from the upper Ediacaran Miaohe Member in South China. Journal of Systematic *Palaeontology*, DOI: 10.1080/14772019.2017.1404499

ŻELAŹNIEWICZ, A., SEGHEDI, A., JACHOWICZ, M., BOBIŃSKI, W., BUŁA, Z. & CWOJDZINSKI, S., 2001. – U-Pb SHRIMP data confirm the presence of a Vendian foreland flysch basin next to the East European Craton. Abstracts, EUROPROBE Conference, Ankara, Turkey, 98-101.

ŻELAŹNIEWICZ A., BUŁA Z., FANNING M., SEGHEDI A. & ŻABA J. 2009. – More evidence on Neoproterozoic terranes in Southern Poland and southeastern Romania. *Geological Quarterly* 53 (1), 93-124.

ZHAO Y., CHEN M., PENG J., YU M., HE M., WANG Y., YANG R., WANG P. & ZHANG Z. 2004. – Discovery of a Miaohe-type Biota from the Neoproterozoic Doushantuo Formation in Jiangkou County, Guizhou Province, China. *Chinese Science Bulletin* 49 (20): 2224-2226.

ZHURAVLEV A.Y. 1993. – Were Ediacaran Vendobionta multicellulars? *Neues Jahrbuch für Geologie und Paläontologie Abhandlungen* 190: 299-314.

**FIGURE CAPTIONS**

**Figure 1**

Schematic geologic map of outcropping Neoproterozoic in Central Dobrogea area and situation of the sampling of studied specimens.

**Figure 2**

*Beltanelliformis brunsae* Menner in Keller et al., 1974 from Dobrogea. **a**. General view of specimen 1; **b**. Detail of the discoid imprints of specimen 1 exhibiting the characteristic small wrinkles at the periphery (black arrows). Note some soft deformities of some individuals (white arrow); **c.** General view of specimen 2; **d**. Detail of the discoid imprints of the specimen 2 (MNHN.F.A68682) exhibiting the characteristic fine wrinkles at the periphery (black arrows). Some individuals may partially cover others (white arrow). Scale-bar: 1 cm.

**Figure 3**

**a**. Thin section across the surface bearing *Beltanelliformis brunsae* imprints of specimen 2. The location of an imprint is indicated by the arrow. The imprint is covered and sealed with a thin detrital bed; **b**. Polished surface across the sediment covering the imprints with coarser detrital supply structured in micro-HCS (arrow). Scale bar: 1 mm for Fig. 3a, 1cm for Fig. 3b.

**Figure 4**

Surface measurements of individual discoid imprints for the two specimens.

**Figure 1**

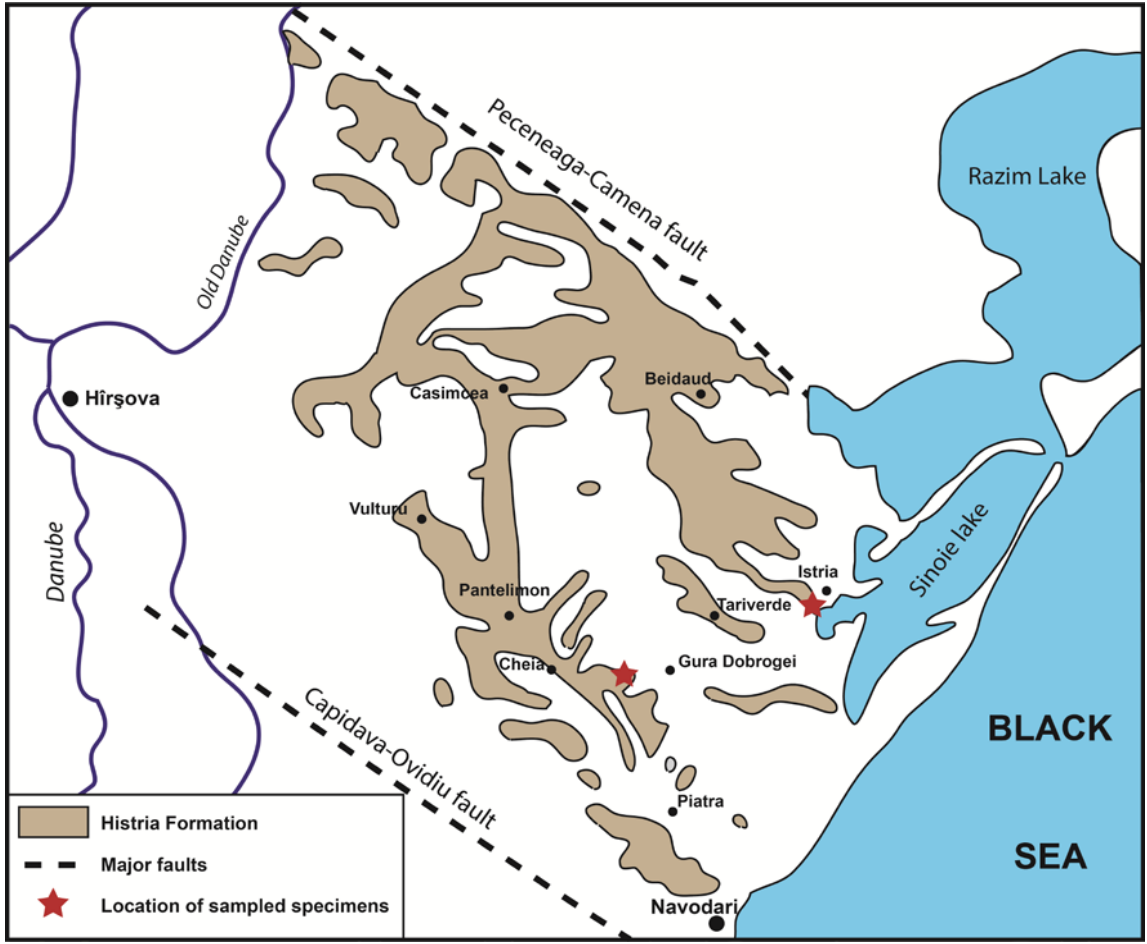

**Figure 2**

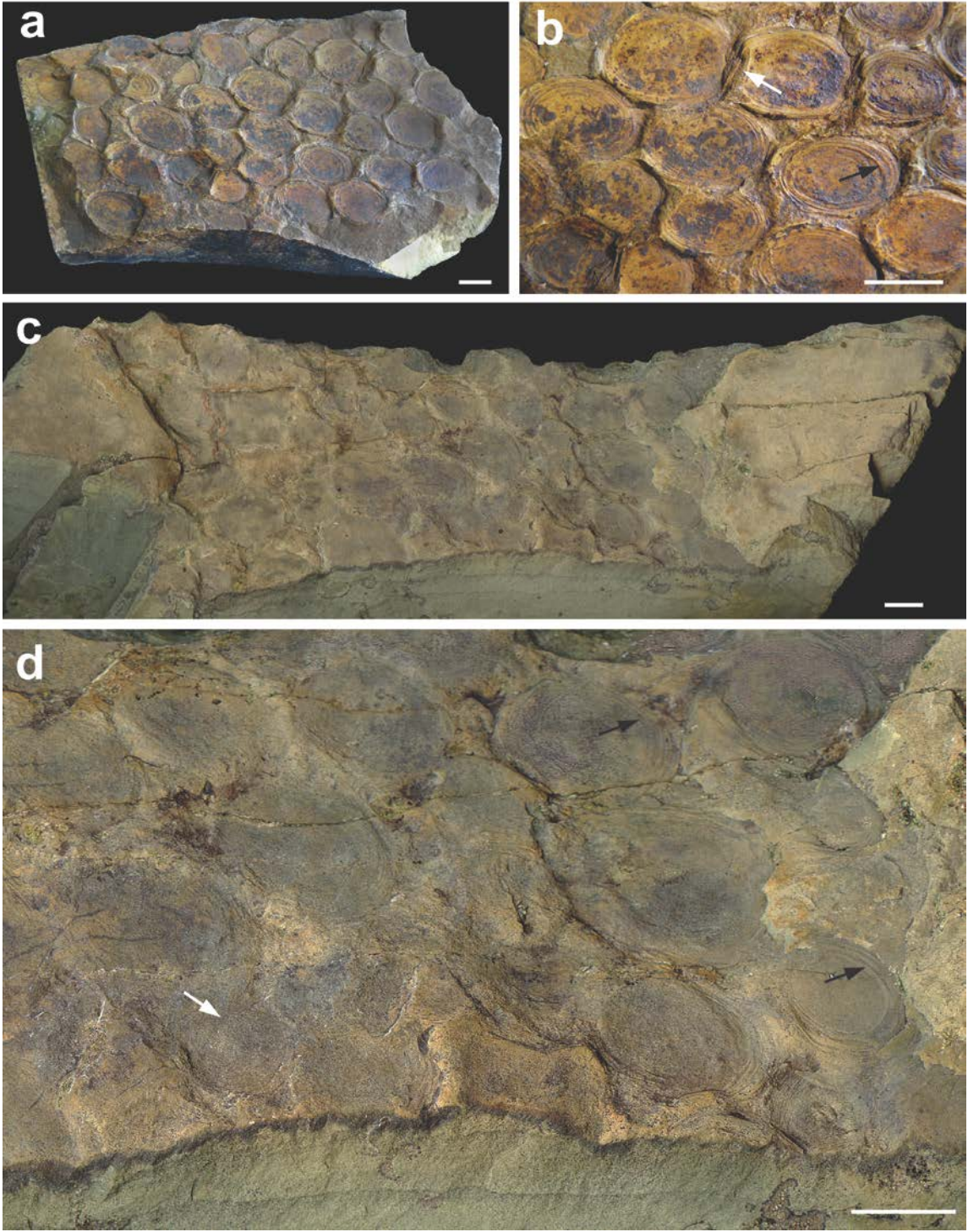

**Figure 3**

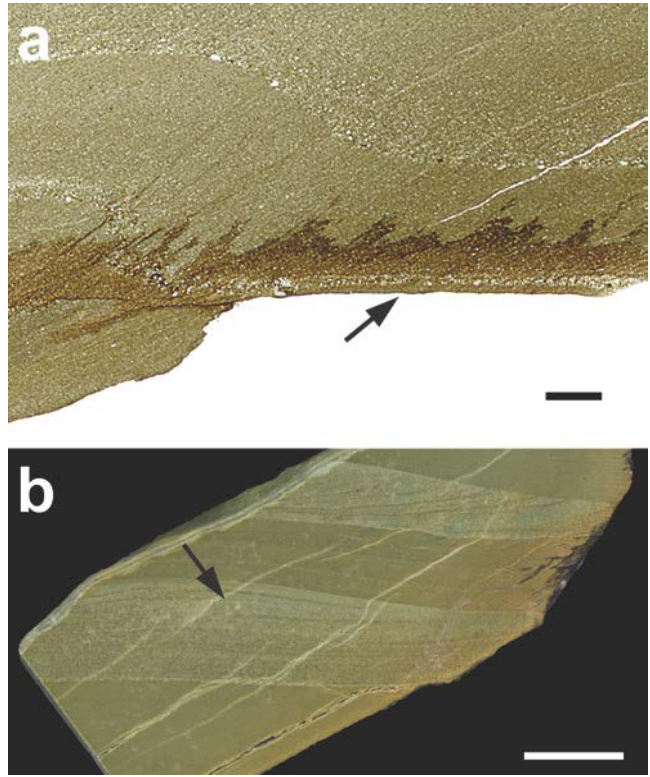

**Figure 4**

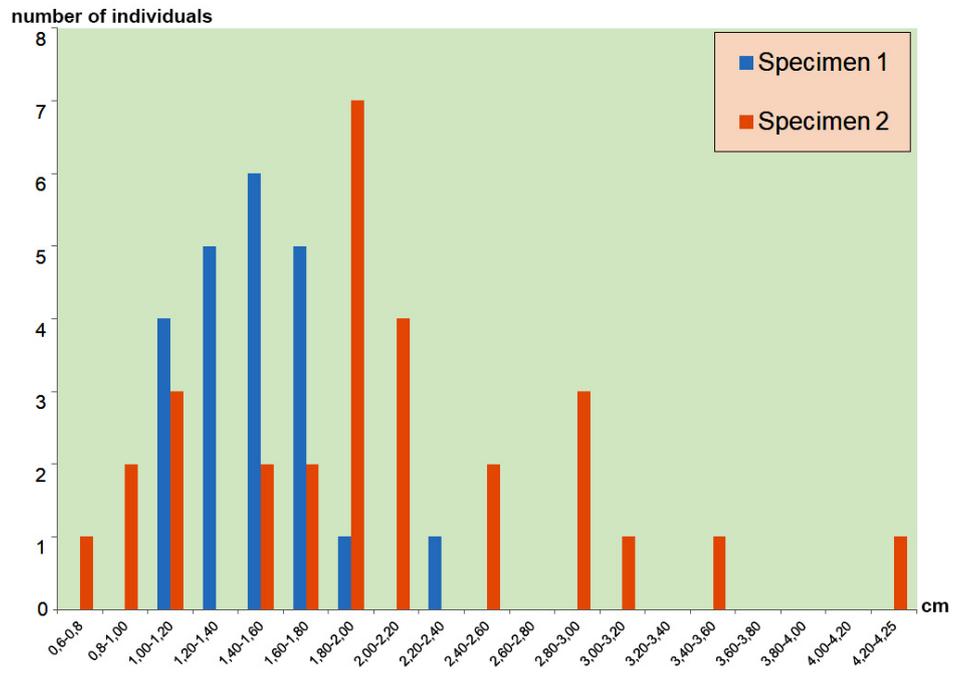